\documentstyle[aps]{revtex}
\def\Journal#1#2#3#4{{#1} {\bf #2}, #3 (#4)}
\input{epsfig.sty}

\def\NPB{{\rm Nucl. Phys.} B}
\def\PLB{{\rm Phys. Lett.}  B}
\def\PRL{\rm Phys. Rev. Lett.}
\def\PRD{{\rm Phys. Rev.} D}
\def\PRC{{\rm Phys. Rev.} C}

\def\be{\begin{equation}}
\def\ee{\end{equation}}
\def\bea{\begin{eqnarray}}
\def\eea{\end{eqnarray}}
\def\lsim{\mathrel{\rlap{\lower4pt\hbox{\hskip1pt$\sim$}}
    \raise1pt\hbox{$<$}}}         
\def\gsim{\mathrel{\rlap{\lower4pt\hbox{\hskip1pt$\sim$}}
    \raise1pt\hbox{$>$}}}
\begin{document}
\draft
\title{Continuity of generalized parton distributions
for the pion virtual Compton scattering}
\author{ Ho-Meoyng Choi \\
 Department of Physics, Carnegie-Mellon University,
Pittsburgh, PA 15213\\
Chueng-Ryong Ji \\
Department of Physics, North Carolina State University,
Raleigh, NC 27695-8202\\
L. S. Kisslinger\\
 Department of Physics, Carnegie-Mellon University,
Pittsburgh, PA 15213 }

\maketitle
\begin{abstract}
We discuss a consistent treatment of the light-front gauge-boson
and meson wave functions in the analyses of
the generalized parton distributions(GPDs) and the scattering amplitudes in
deeply virtual Compton scattering(DVCS) for the pion.
The continuity of the GPDs at the crossover, where
the longitudinal momentum fraction of the probed quark
is same with the skewedness parameter, and the finiteness of the DVCS
amplitude are ensured if the same light-front radial wave function as
that of the meson bound state wave function is used for
the gauge boson bound state arising from the pair-creation(or
nonvalence) diagram.
The frame-independence of our model calculation is also guaranteed
by the constraint from the sum rule between the GPDs and the form factors.
\end{abstract}
PACS number(s): 13.40.Gp, 12.39.Ki, 13.60.Fz
\date{}
\section{Introduction}
Compton scattering provides a unique tool
for studying hadronic structure. The Compton amplitude probes the hadrons
through a coupling of two electromagnetic currents and thus can be regarded
as a generalization of hadron form factors.
For the Compton amplitude, when the initial photon is highly virtual but the
final one is real, one arrives at the kinematics of deeply virtual
Compton scattering(DVCS). According to the Quantum ChromoDynamics(QCD),
the amplitudes at large momentum transfer factorize in the form of a
convolution of a hard scattering amplitude which can be computed
perturbatively from quark-gluon subprocesses multiplied by
process-independent distribution amplitudes containing the
bound-state nonperturbative dynamics for each of the interacting hadrons.
Thus, the most important contribution to DVCS amplitude is given
by the convolution of a hard quark propagator and a nonperturbative function
describing long-distance dynamics which is known as ``generalized parton
distributions(GPDs)" and also known as ``skewed parton
distributions(SPDs)"~\cite{Mu,XJ1,Ra1}. The latter has been studied
extensively in searching for possible new physics(see, for example,
Ref.~\cite{GPV} and references therein).

The GPDs serve as a generalization of the ordinary (forward) parton
distributions and provide much more direct and sensitive information on
the light-front(LF) wave function of a target hadron than the hadron form
factors. In particular, the momentum of the ``probed quark" in GPDs is not
integrated over, but rather kept fixed at longitudinal
momentum fraction $x$, while for the form factor it is integrated out
(due to the nonlocal current operator
$\bar{\psi}(0)\gamma^\mu\psi(z)$ for the GPDs in contrast to
the local vertex $\bar{\psi}(0)\gamma^\mu\psi(0)$ for the form factor).
The form factors are then just moments of the GPDs.
At the expense of being generalized amplitudes, the GPDs always involve
the nonvalence contributions due to the nature of longitudinal asymmetry
$\zeta=(P-P')^+/P^+$(so called ``skewedness" parameter) between
initial($P$) and final($P'$) hadron state momenta.
While the kinematic region where the longitudinal momentum fraction
$x$ of the probed quark is greater than the skewedness parameter $\zeta$
(i.e. $1>x>\zeta$) is called ``DGLAP region"\cite{DGLAP}, the rest
of the longitudinal momentum region $0<x<\zeta$ is called
``ERBL region"\cite{ERBL}.
The DGLAP and ERBL regions have also been denoted as the valence and
nonvalence regions in the LF dynamics, respectively, because the
parton-number-changing nonvalence Fock-state contributions cannot be
avoided for $0<x<\zeta$ while only the parton-number-conserving valence
Fock-state contributions are needed for $1>x>\zeta$. Thus, it has been a
great challenge to calculate the
nonvalence contributions to the GPDs in the framework of LF quantization.

Although many recent theoretical endeavors~\cite{Dh1,Dh2,BDH,Bur1,TM,CJK_SPD}
have been made in describing the GPDs in terms of LF wave
functions, the task has not yet been satisfactory enough for
practical calculations. In Refs.~\cite{Dh2} and \cite{BDH}, the nonvalence
contributions to the GPDs have been rewritten in terms of LF
wave functions with different parton configurations. However, the
representation given in Refs.~\cite{Dh2} and \cite{BDH} requires to find all
the higher Fock-state wave functions while there has been relatively little
progress in computing the basic wave functions of hadrons from first
principles. In Refs.~\cite{Bur1} and \cite{TM}, the GPDs were expressed in
terms of LF wave function but only within toy models such as
the 't~Hooft model of $(1+1)$-dimensional QCD~\cite{Bur1} and the scalar
Wick-Cutkosky model~\cite{TM}, respectively.
While these toy model analyses are helpful
to gain some physical insight on the properties of the GPDs (especially,
the time reversal invariance, the continuity at the crossover
between the DGLAP and ERBL regions, and the sum rule constrained
by the electromagnetic form factor), the real $(3+1)$-dimensional QCD
motivates us to come up with the more realistic model for the application
to the analysis of GPDs.

In an effort toward this direction, we have presented an effective
treatment of handling the nonvalence contributions to the GPDs of the pion
\cite{CJK_SPD} using our LF constituent quark model(LFQM),
which has been phenomenologically quite successful in describing
the spacelike form factors for the electromagnetic and radiative decays of
pseudoscalar and vector mesons~\cite{CJ1,CJ2,KCJ} and the
timelike weak form factors for exclusive semileptonic and
rare decays of pseudoscalar mesons~\cite{JC,Time,Rare}.
Our effective treatment of handling the nonvalence contributions is based
on the covariant Bethe-Salpeter(BS) approach formulated in the LF
quantization\cite{JC} which we call LFBS approach and has been previously
applied to the exclusive semileptonic and rare decays of pseudoscalar
mesons~\cite{Time,Rare} providing reasonable results compared to the data.

However, an artifact of discontinuity at $x=\zeta$ occured in our previous
calculation of GPDs\cite{CJK_SPD}.
As we provided the reasoning\cite{CJK_SPD}, the discontinuity is caused
by the different $x=\zeta$ behavior between the gauge boson vertex and the
hadronic vertex if the wave function for the gauge boson vertex is taken
differently from that for the hadronic vertex.
The similar observation was made recently in Ref.\cite{TM} using a
different model(scalar Wick-Cutkosky model).
Such discontinuity at $x=\zeta$ may cause a divergence in the DVCS amplitude.

In this work, we improve our previous analysis\cite{CJK_SPD} by taking
the same LFBS approach for both vertices of meson and gauge boson and
explicitly show that our reasoning presented in Ref.\cite{CJK_SPD} is
correct; i.e. the continuity of GPDs at the crossover($x=\zeta$) is
ensured by this consistent treatment of vertices. We also
discuss the value of GPDs at $x=\zeta$ in conjunction with the single
spin asymmetry (SSA)\cite{KPV,FM} and calculate the scattering amplitude
contributing
dominantly to the Compton scattering of the pion in the deeply virtual
region.
The paper is organized as follows. In Section II, we derive the GPDs
as a nonperturbative formulation of the light-front dominated
deeply virtual Compton scattering(DVCS) $\gamma^*\pi\to\gamma\pi$ of the
pion and introduce the necessary kinematics following the notation
employed by Radyushkin~\cite{Ra1}.
In Section III, we discuss the LFBS approach and present the
consistent treatment of the gauge boson and hadron vertex functions
in handling the nonvalence contributions to the GPDs.
The implication of the vector meson dominance(VMD) at the gauge boson
vertex in this improved analysis and the GPD value at $x=\zeta$ are also
discussed. In Section IV, we show our numerical results for the GPDs of
the pion that satisfy the continuity at the crossover and in turn give the
finite scattering amplitude for DVCS of the pion. The frame-independence
of our model is checked by the sum rule between the GPDs and the pion
form factor. We also comment on the polynomiality conditions
associated with the D-term contribution\cite{PW,KPV} in ERBL region.
Conclusions follow in Section V.

\section{Scattering Amplitude in DVCS}

\begin{figure}
\centerline{\psfig{figure=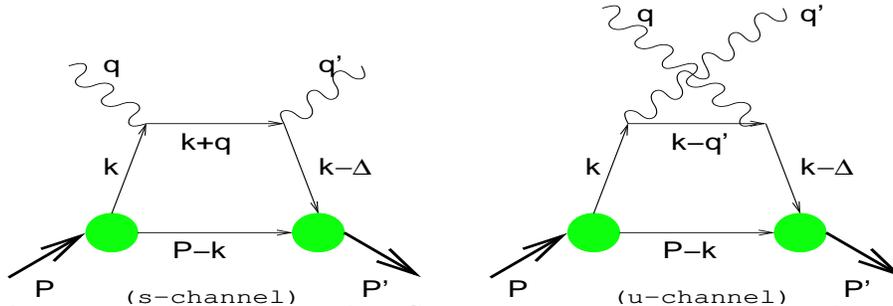,height=4cm,width=12cm}}
\caption{Handbag diagrams contributing dominantly to Compton scattering in
the deeply virtual region. The lower soft part consists of a hadronic matrix
element which is parametrized in the form of generalized parton distribution
functions.\label{handbag}}
\end{figure}

We begin with the kinematics of the virtual Compton
scattering (see Fig.~\ref{handbag}) of the pion
\be\label{VCS}
\gamma^*(q) + \pi(P) \to \gamma(q') + \pi(P'),
\ee
where the initial (final) hadron state is characterized by the momentum
$P\;(P')$ and the incoming spacelike virtual and outgoing real photon
momenta by $q$ and $q'$, respectively\footnote{If both photons are far
off-shell and have equal spacelike virtuality, it becomes virtual forward
Compton amplitude and its imaginary part determines structure functions
of deep inelastic scattering(DIS). If they are both real but the momentum
transfer is large, it becomes wide-angle Compton scattering(WACS)
amplitude~\cite{WACS}.}. We shall use the component notation
$V=(V^+,V^-,{\bf V}_\perp)$ and our metric is specified by
$V^{\pm}=(V^0 \pm V^3)$ and $V \cdot V=V^+V^- - {\bf V}^2_{\perp}$.

Defining the four momentum transfer $\Delta=P-P'$, one has
$P =[P^+, M^2/P^+, 0_\perp]$,
$P'=[(1-\zeta)P^+, (M^2+\Delta^{2}_{\perp})/(1-\zeta)P^+, -\Delta_\perp]$,
and
$\Delta = P - P' =[\zeta P^+, (\Delta^2+\Delta^2_\perp)/\zeta P^+, \Delta_\perp]$,
where $M$ is the pion mass and $\zeta=\Delta^+/P^+$ is
the skewedness parameter describing the asymmetry in plus momentum.
The squared momentum transfer then reads
\be\label{del2}
t = \Delta^2 = 2 P\cdot\Delta=-\frac{\zeta^2 M^2 + \Delta^2_\perp}{1-\zeta}.
\ee
Since $\Delta^2_\perp\geq 0$, $t$ has a minimum value
$-t_{\min}=\zeta^2 M^2/(1-\zeta)$ at given $\zeta$.
As shown in Fig.~\ref{handbag}, the parton emitted by the pion
has the momentum $k$, and the one absorbed has the
momentum $k'=k-\Delta$.
As in the case of spacelike form factors, we choose a frame
where the incident spacelike photon carries $q^+=0$:
$q = [0, ({\bf q}_\perp + \Delta_\perp)^2/\zeta P^+
+ (\zeta M^2 +\Delta^2_\perp)/(1-\zeta) P^+, {\bf q}_\perp$] and
$q' = [\zeta P^+, ({\bf q}_\perp + \Delta_\perp)^2/\zeta P^+,
{\bf q}_\perp +\Delta_\perp]$.

In deeply virtual Compton scattering (DVCS) where $Q^2=-q^2$ is large
compared to the mass $M$ and $-t$, one obtains
$Q^2/(2P\cdot q)=\zeta$, i.e. $\zeta$ plays the role of the
Bjorken variable in DVCS. For a fixed
value of $-t$, the allowed range of $\zeta$ is given by
\be\label{range}
0\leq\zeta\leq\frac{(-t)}{2M^2}\biggl(
\sqrt{1 + \frac{4M^2}{(-t)}}-1\biggr).
\ee

In the leading twist ignoring interactions at the
quark-gauge boson(photon in this case) vertex,
the amplitude contributing dominantly to Compton scattering
in the deeply virtual region is given by

\bea\label{Mmn}
M^{\mu\nu}=M^{\mu\nu}_s + M^{\mu\nu}_u,
\eea
where $M^{\mu\nu}_s$ and $M^{\mu\nu}_u$ are the $s(=(P+q)^2)$- and
$u(=(P-q')^2)$-channel amplitudes(see Fig.~\ref{handbag}), respectively.
The $s$-channel amplitude is given by
\bea\label{Mmns}
M^{\mu\nu}_s&=& -iN_ce^2_q\int\frac{d^4k}{(2\pi)^4}
{\rm Tr}[\gamma_5(\not\!k +m)\gamma^\mu(\not\!k+\not\!q+m)\gamma^\nu
(\not\!k-\not\!\Delta+m)\gamma_5(-\not\!P+\not\!k+m)]
\nonumber\\
&\times&
\frac{H_{\rm cov}(k,P)H'_{\rm cov}(k-\Delta, P-\Delta)}
{[k^2-m^2+i\varepsilon][(k+q)^2-m^2+i\varepsilon]
[(k-\Delta)^2-m^2+i\varepsilon][(P-k)^2-m^2+i\varepsilon]},
\eea
where $N_c$ is the color factor and
$H_{\rm cov}(k,P)[H'_{\rm cov}(k-\Delta,P-\Delta)]$ is the covariant
initial[final] state meson-quark vertex function that satisfies the
BS equation. As usual in the LFBS formalism,
we assume that the covariant vertex function $H_{\rm cov}(k)$
does not alter the $k^-$ pole structure in Eq.(\ref{Mmns}).
The $u$-channel amplitude can be easily obtained by
$M^{\mu\nu}_u=M^{\mu\nu}_s(q\to-q')$.

In deeply virtual $q^-\simeq Q^2/\zeta P^+$ limit, one obtains
$(\not\!k+\not\!q+m)\simeq\not\!q\simeq\gamma^+q^-/2$
in the trace term of Eq.~(\ref{Mmns}). Consequently, combining this
with the 2nd term of the denominator in Eq.~(\ref{Mmns}) leads to
$q^-/[(k+q)^2-m^2+i\varepsilon]\simeq 1/[P^+(x-\zeta+i\varepsilon)]$,
where $x=k^+/P^+$.  Similarly, one can obtain
$(-q')^-/[(k-q')^2-m^2+i\varepsilon]\simeq 1/[P^+(x-i\varepsilon)]$
from the $u$-channel amplitude.

Adding these two $s$- and $u$-channel amplitudes,
we obtain the Compton scattering amplitude in DVCS limit as follows
\bea\label{CDVCS}
M^{IJ}= \epsilon^I_\mu\epsilon^{*J}_\nu M^{\mu\nu}
&=&-\frac{iN_c}{2P^+}e^2_q\int\frac{d^4k}{(2\pi)^4}
\biggl(\frac{1}{x-\zeta+i\varepsilon}+
\frac{1}{x-i\varepsilon}\biggr)
H_{\rm cov}(k,P)H'_{\rm cov}(k-\Delta, P-\Delta)
\nonumber\\
&\times&
\frac{
{\rm Tr}[\gamma_5(\not\!k +m)
\not\!\epsilon^I\gamma^+\not\!\epsilon^{*J}
(\not\!k-\not\!\Delta+m)\gamma_5(-\not\!P+\not\!k+m)] }
{[k^2-m^2+i\varepsilon] [(k-\Delta)^2-m^2+i\varepsilon]
[(P-k)^2-m^2+i\varepsilon]}.
\eea
For circularly polarized($\epsilon^+=0$) initial and final
photons\footnote{As discussed in~\cite{BDH}, for a longitudinally
polarized initial photon, the Compton amplitude is of order $1/Q$
and thus vanishes in the limit $Q^2\to\infty$.}($I,J$ are $\uparrow$
or $\downarrow$),
we obtain from Eq.~(\ref{CDVCS})
\bea\label{pol}
\not\!\epsilon^I\gamma^+\not\!\epsilon^{*J}
&=& (\epsilon^I_\perp\cdot\epsilon^{*J}_\perp)\gamma^+
+ i(\epsilon^I_\perp\times\epsilon^{*J}_\perp)_3\gamma^+\gamma_5,
\eea
where we use the identities $(\gamma^+)^2=0$,
$\gamma^1\gamma^+\gamma^1=\gamma^2\gamma^+\gamma^2=\gamma^+$ and
$\gamma^1\gamma^+\gamma^2=-\gamma^2\gamma^+\gamma^1=i\gamma^+\gamma_5$.
Equation~(\ref{pol}) reduces to $\gamma^+(1\pm\gamma_5)$ for
the parallel helicities(i.e. $+$ for $\uparrow\uparrow$ and
$-$ for $\downarrow\downarrow$) and zero otherwise.
Since the axial current $\gamma^+\gamma_5$ does not contribute
to the integral, i.e. $\sim({\bf k}_\perp\times\Delta_\perp)$ after
the trace calculation, we shall omit this term from now on.

Then, the DVCS amplitude(i.e. photon helicity amplitude)
can be rewritten as the factorized form of hard and soft amplitude
\bea\label{CDVCS2}
M^{\uparrow\uparrow}(P,q,P') = M^{\downarrow\downarrow}(P,q,P')
&=&-e^2_q\int\; dx \biggl(\frac{1}{x-\zeta+i\varepsilon}+
\frac{1}{x-i\varepsilon}\biggr){\cal F}_\pi(\zeta,x,t),
\eea
where
\bea\label{HPI}
{\cal F}_\pi(\zeta,x, t)=
&=&\frac{iN_c}{2}\int\frac{dk^-d^2{\bf k}_\perp}{2(2\pi)^4}
H_{\rm cov}(k,P)H'_{\rm cov}(k-\Delta, P-\Delta)
\nonumber\\
&\times&
\frac{
{\rm Tr}[\gamma_5(\not\!k +m)\gamma^+
(\not\!k-\not\!\Delta+m)\gamma_5(-\not\!P+\not\!k+m)] }
{[k^2-m^2+i\varepsilon] [(k-\Delta)^2-m^2+i\varepsilon]
[(P-k)^2-m^2+i\varepsilon]}.
\eea
The function ${\cal F}_\pi(\zeta,x,t)$ is so called
``generalized parton distributions" and it manifests characteristics of
the ordinary(forward) quark distribution in the limit of $\zeta\to 0$
and $t\to 0$. On the other hand,  the first moment of the
${\cal F}_\pi(\zeta,x,t)$ \footnote{Note that our definition
of ${\cal F}_\pi(\zeta,x,t)$ in this work is different from the
one used in our previous work \cite{CJK_SPD} by a
factor $(1-\zeta/2)$ in the normalization. We also used the notation
$\zeta$ for the "skewedness" parameter instead of previous $\xi$
to be consistent with the notation used by Radyushkin \cite{Ra1}.}
is related to the form factor by the
following sum rules~\cite{XJ1,Ra1}:
\be\label{sum}
\int^1_{0}\frac{dx}{1-\frac{\zeta}{2}}
{\cal F}_{\pi}(\zeta, x, t)
=F_\pi(t).
\ee
In general, the polynomiality conditions for the moments of
the GPDs\cite{Jsong,Ra2} defined by
\be\label{sum-mom}
\int^1_{0}\frac{dx}{1-\frac{\zeta}{2}} x^{n-1} {\cal F}_{\pi}(\zeta, x, t)
=F_n(\zeta,t)
\ee
require that the highest power of $\zeta$ in the polynomial expression of
$F_n(\zeta,t)$ should not be larger than $n$. These polynomiality
conditions are fundamental properties of the GPDs which follow from the
Lorentz invariance. We comment on how our model calculations satisfy the
polynomiality conditions in Section IV(numerical results).

An important feature of the DVCS amplitude given by Eq.~(\ref{CDVCS2})
is that it depends only on the skewedness parameter $\zeta=Q^2/(2P\cdot q)$
for large $Q^2$ and fixed $|t|(\leq Q^2)$, i.e. DVCS is equivalent to an
inclusive process exhibiting the Bjorken scaling $x_{Bj}=\zeta$.
Note also from Eq.~(\ref{CDVCS2}) that the imaginary part of the DVCS
amplitude is proportional to ${\cal F}_\pi(\zeta,\zeta,t)$.
The single spin asymmetry(SSA)\cite{KPV,FM} that can be measured in the
scattering of a longitudinally polarized probe on an unpolarized target is
proportional to the imaginary part of the amplitude, {\it i.e.}
the value of ${\cal F}_\pi (\zeta,\zeta,t)$.
The recent measurements of SSA for the
proton target have been reported by HERMES \cite{HERMES} and CLAS \cite{CLAS}
collaborations. We discuss the value of ${\cal F}_\pi(\zeta,\zeta,t)$
in Section III C.

\section{ Generalized parton distributions in light-front Bethe-Salpeter
approach}
\begin{figure}
\centerline{\psfig{figure=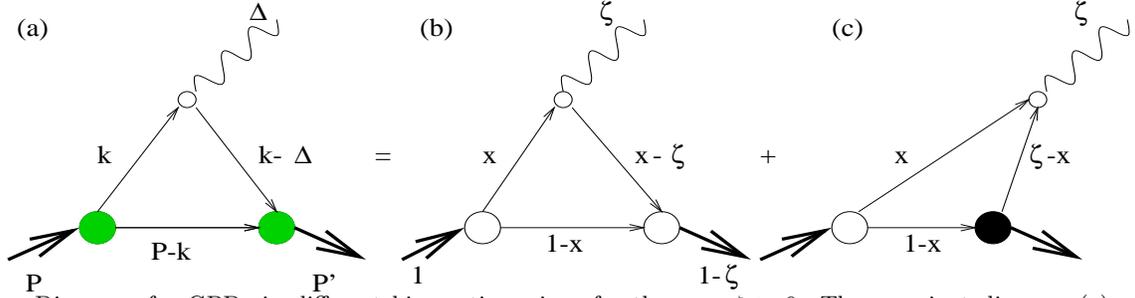,height=4cm,width=15cm}}
\caption{ Diagrams for GPDs in different kinematic regions for the
case $\zeta>0$: The covariant diagram (a) corresponds to the sum of
the LF valence diagram (b) defined in DGLAP($\zeta< x<1$) region
and the nonvalence diagram (c) defined in ERBL($0< x<\zeta$) region.
The large white and black blobs at the meson-quark vertices in (b) and
(c) represent the ordinary LF wave function and the nonvalence
wave function vertices, respectively.
The small white blob at the quark-gauge boson
vertex indicates the nonlocality of the vertex.
\label{highFock}}
\end{figure}

The GPDs entering as nonperturbative information in DVCS as shown in
Fig.~\ref{handbag} can be represented by Fig.~\ref{highFock}(a) where the
small white blob shown in Fig.~\ref{highFock}(a)
represents the composite(nonlocal) operator~\cite{Ra1}
at the quark-gauge boson vertex.
Since the longitudinal component, $\Delta^+=\zeta P^+$, of the momentum
transfer $\Delta=P-P'$ is in general nonzero, the covariant diagram
Fig.~\ref{highFock}(a) is represented by the sum of
the LF valence diagram (b) defined in $\zeta< x<1$ (DGLAP) region
and the nonvalence diagram (c) defined in $0< x<\zeta$ (ERBL) region.
As one can see from Fig.~\ref{highFock}(b) and~(c), the large white
blobs at the meson-quark vertices in (b) and
(c) represent the ordinary LF wave function. However, the large black
blob in (c) cannot be represented by the ordinary LF wave function since it
is no longer a bound state vertex but non-wave-function vertex.
This non-wave-function vertex causes the main source of difficulty
in representing the GPDs(as well as the timelike form factor) in terms
of light-front wave function.
However, in a covariant BS formalism it can be represented as an
analytic continuation of the usual BS amplitude.

In our previous work~\cite{CJK_SPD}, we have derived the GPDs of the pion
starting from the covariant BS amplitude of the current
${\cal J}$(see Eq.~(12) in~\cite{CJK_SPD}),
which is essentially the same calculation of ${\cal F}_\pi(\zeta,x,t)$ given
by Eq.~(\ref{HPI}).
Since the detailed procedures for obtaining the effective solution for
the non-wave-function vertex have been given in~\cite{CJK_SPD}, here
we only briefly present the salient points of our previous effective
method~\cite{CJK_SPD} before we discuss the new treatment of the quark-gauge
boson vertex function to ensure the continuity of the GPDs
at the crossover($x=\zeta$) between the DGLAP and ERBL regions.

\subsection{Brief review of our previous method~\protect\cite{CJK_SPD}}

\begin{figure}
\centerline{\psfig{figure=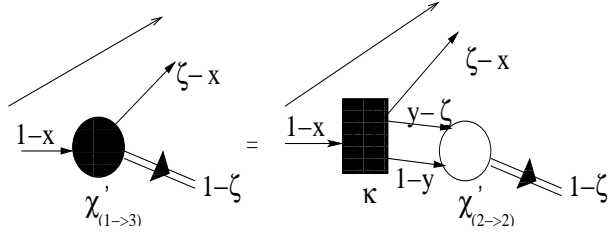,height=3cm,width=8cm}}
\caption{Non-wave-function vertex(black blob) linked to an ordinary
light-front wave function(white blob).\label{SDFig}}
\end{figure}

The essential feature of our approach is to consider the light-front
wave function as the solution of light-front Bethe-Salpeter
equation(LFBSE) given by~\cite{CJK_SPD,JC,BJS,Sales}
\bea\label{SDtype1}
(M^2- M'^2_0)\chi'(x'_i,{\bf k'}_{i\perp})
&=&\int^1_0 [dy'][d^2{\bf l'}_\perp]
{\cal K}(x'_i,{\bf k'}_{i\perp}; y'_j,{\bf l'}_{j\perp})
\chi'(y'_j,{\bf l'}_{j\perp}),
\eea
where\footnote{We present the LFBSE for the final state
meson-quark vertex with the final state momentum variables. However,
the same form of LFBSE can be used in the initial state meson-quark
vertex.} ${\cal K}$ is the BS kernel which in principle includes all the
higher Fock-state contributions, $M'^2_0=(m^2+{\bf k'}^2_{\perp})/x'
+ (m^2 + {\bf k'}^2_{\perp})/(1-x')$ is the invariant mass, and
$\chi'(x'_i,{\bf k'}_{i\perp})$ is the (final state) BS amplitude with
the internal momenta of the (struck) quark for the final state,
$x'=(x-\zeta)/(1-\zeta)$ and ${\bf k'}_\perp={\bf k}_\perp + x'\Delta_\perp$.
The internal momenta of the (struck) quark after the kernel are given
by $y'=(y-\zeta)/(1-\zeta)$ and ${\bf l'}_\perp={\bf l}_\perp + y'\Delta_\perp$
so that the integration of $y'$ runs from 0 to 1(or $y$ runs from $\zeta$ to 1).
Note also  $d^2{\bf l'}_\perp = d^2{\bf l}_\perp$ for a given $y'$.
We define the valence BS amplitude(i.e. $x > \zeta$)
as $\chi_{(2\to 2)}$ and the nonvalence BS amplitude(i.e. $x < \zeta$) as
$\chi_{(1\to 3)}$ where the subscript indicates the parton number
before and after the kernel.
Both the valence and nonvalence BS amplitudes are solutions
to Eq.~(\ref{SDtype1}).
As illustrated in Fig.~\ref{highFock}(c),
the nonvalence BS amplitude is an analytic continuation of the
valence BS amplitude. In the LFQM the relationship between the
BS amplitudes in the two regions is given by~\cite{CJK_SPD,JC}
\bea\label{SDtype2}
(M^2-M'^2_0)\chi'_{(1\to 3)}(x'_i,{\bf k'}_{i\perp})
&=&\int^1_0 [dy'][d^2{\bf l}_\perp]
{\cal K}(x'_i,{\bf k'}_{i\perp}; y'_j,{\bf l'}_{j\perp})
\chi'_{(2\to 2)}(y'_j,{\bf l'}_{j\perp}),
\eea
where again the kernel includes in principle all the higher Fock-state
contributions because all the higher Fock components of the bound-state are
ultimately related to the lowest Fock component with the use of kernel as
illustrated in Fig.~\ref{SDFig}.

After the Cauchy integration over $k^-$ in Eq.~(\ref{HPI}),
we obtain the valence($\zeta\leq x\leq 1$) and
nonvalence($0\leq x\leq\zeta $) contributions to the GPDs
of the pion as follows
\bea\label{jv}
{\cal F}^{val}_\pi(\zeta,x,t)&=& \frac{N_c}{2x(1-x)x'}
\int \frac{d^2{\bf k}_\perp}{16\pi^3}
\chi_{(2\to 2)}(x,{\bf k}_\perp)S^+_{val}(x,{\bf k}_\perp)
\chi'_{(2\to 2)}(x',{\bf k}'_{\perp}),
\eea
and
\bea\label{jnv}
{\cal F}^{nv}_\pi(\zeta,x,t)
&=&\frac{-N_c}{2x(1-x)(1-x'')}\int
\frac{d^2{\bf k}_\perp}{16\pi^3}
\chi_{(2\to 2)}(x,{\bf k}_\perp)S^+_{nv}(x,{\bf k}_\perp)
\chi^g(x'',{\bf k''}_\perp)
\nonumber\\
&\times&\int^1_0\frac{dy'}{y'(1-y')}\int d^2{\bf l}_\perp
{\cal K}(x',{\bf k'}_\perp; y',{\bf l'}_\perp)
\chi'_{(2\to 2)}(y',{\bf l'}_\perp),
\eea
where Eq.~(\ref{SDtype2}) has been used for the non-wave-function
vertex as shown in Fig.~\ref{SDFig}[or Fig.~\ref{highFock}(c)].
The trace terms of the quark propagators, i.e.
$S^+_{val}(x,{\bf k}_\perp)$ and $S^+_{nv}(x,{\bf k}_\perp)$,
are given by Eqs.~(21) and (25) in Ref.~\cite{CJK_SPD}, respectively.
The light-front vertex functions of the initial[final]
hadron $\chi_{(2\to2)}[\chi'_{(2\to2)}]$ and the gauge boson
$\chi^g$ are given by\footnote{Note that we slightly modified the
definition of $\chi^g$ in present work from the one in~\cite{CJK_SPD},
i.e. we take the common factor $\zeta/(1-\zeta)$ from Eq.~(24) in
Ref.~\cite{CJK_SPD} and put back into the prefactor leading to the
present Eq.~(\ref{jnv}).}
\bea\label{wfI}
\chi_{(2\to 2)}(x,{\bf k}_\perp)&=&\frac{h_{LF}}{M^2 -M^2_0},\;
\chi'_{(2\to 2)}(x',{\bf k'}_\perp)=\frac{h'_{LF}}{M^2 -M'^2_0},\;
\chi^g(x'',{\bf k''}_\perp)=\frac{1}{\Delta^2-M''^2_0},
\eea
where $x''=x/\zeta$, ${\bf k''}_\perp={\bf k}_\perp + x''\Delta_\perp$,
and $M''^2_{0}=({\bf k''}^2_\perp + m^2)/x''
+ ({\bf k''}^2_\perp + m^2)/(1-x'')$ is the invariant mass of $q\bar{q}$
pair annihilated into the gauge boson.
Equations~(\ref{jv}) and~(\ref{jnv}) were our essential results from
our previous analysis~\cite{CJK_SPD}.
Since the Eqs.~(\ref{jv}) and~(\ref{jnv}) are divergent themselves
with a constant LF vertex function $h_{LF}$, our idea is to replace
the $h_{LF}$[or equivalently $\chi_{(2\to2)}(x,{\bf k}_\perp$)] with
the standard LF vertex
function(i.e. dressed vertex)~\cite{CJ1,CJ2,JC,Time,Ja},
which has been successful in predicting many static properties
of ground state mesons. However, we left the gauge boson vertex
function $\chi^g$ as the bare vertex given by Eq.(\ref{wfI})
in our previous work~\cite{CJK_SPD}. We(as well as the
authors in~\cite{TM}) attributed this different treatment between
gauge boson and meson vertex functions to the reason for the artifact of the
discontinuity in the GPDs at the crossover($x=\zeta$) between the
DGLAP and ERBL regions.

\subsection{Consistent treatment of gauge boson and meson vertex
functions in LFQM: Update}

\begin{figure}
\centerline{\psfig{figure=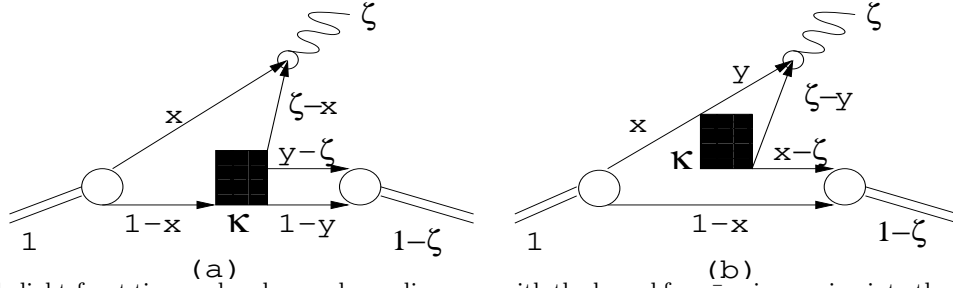,height=1.5in,width=5in}}
\caption{Possible light-front time-ordered nonvalence diagrams
with the kernel for $q\bar{q}$ pair merging into the final state
meson(a) and annihilating into external gauge boson(b).\label{Eff_photon}}
\end{figure}

As discussed in our previous work~\cite{CJK_SPD}
and the works by others~\cite{Ein,Bur1,TM},
in principle one should consider the same BS kernel ${\cal K}$ for
$q{\bar q}$ pair annihilating into external gauge boson as that
for $q\bar{q}$ pair merging into the final state meson given
by Eq.~(\ref{jnv}). This is illustrated in Fig.~\ref{Eff_photon} and
we shall denote the GPDs coming from Figs.~\ref{Eff_photon}(a) and (b)
as ${\cal F}^{nv}_{\pi(a)}(\zeta,x,t)$ (given by Eq.~(\ref{jnv})) and
${\cal F}^{nv}_{\pi(b)}(\zeta,x,t)$, respectively.
Adding both contributions of ${\cal F}^{nv}_{\pi(a)}$ and
${\cal F}^{nv}_{\pi(b)}$ assures
the cancellation of any infrared divergence that might occur in the
kernel ${\cal K}$.

The GPDs from Figs.~\ref{Eff_photon}(a) and (b) can be written
generically(dropping, for simplicity, relative label for internal
momenta)
\bea\label{had-pho1}
\int^\zeta_0 dx {\cal F}^{nv}_{\pi(a)}(\zeta,x,t)
&=&\int^\zeta_0 dx\int d^2{\bf k}_\perp
\int^1_\zeta dy \int d^2{\bf l}_\perp
\chi^{\rm f}(y,{\bf l}_\perp)
{\cal K}_a(x,{\bf k}_\perp; y,{\bf l}_\perp)
\chi^g(x,{\bf k}_\perp)
S^+_{nv}(x,{\bf k}_\perp)
{\tilde\chi}^{\rm i}_a(x,{\bf k}_\perp),
\eea
\bea\label{had-pho2}
\int^\zeta_0 dy {\cal F}^{nv}_{\pi(b)}(\zeta,y,t)
&=&\int^\zeta_0 dy \int d^2{\bf l}_\perp
\int^1_\zeta dx \int d^2{\bf k}_\perp
\chi^{\rm f}(x,{\bf k}_\perp)
{\cal K}_b(y,{\bf l}_\perp; x,{\bf k}_\perp)
\chi^g(y,{\bf l}_\perp)
S^+_{nv}(x,{\bf k}_\perp)
{\tilde\chi}^{\rm i}_b(x,{\bf k}_\perp),
\eea
where the superscript i(f) in $\chi$ indicates the initial(final)
valence light-front wave function(i.e. white blob in
Fig.~\ref{Eff_photon}) and
the notation of ${\tilde\chi}^{\rm i}_{a(b)}$ implies the product of
intitial $\chi^{\rm i}_{a(b)}$ in Fig.~\ref{Eff_photon}a(b)
and other terms(such as prefactor in Eq. (\ref{jnv}))
that depend on the internal momentum variables.
The kernel ${\cal K}_b(y,{\bf l}_\perp;x,{\bf k}_\perp)$ in
Eq.~(\ref{had-pho2}) is equivalent  to
${\cal K}_a(x,{\bf k}_\perp;y,{\bf l}_\perp)$ in Eq.~(\ref{had-pho1})
upon exchange of dummy variables
$(x,{\bf k}_\perp)\leftrightarrow (y,{\bf l}_\perp)$ except an overall
sign due to an exchange of quark and antiquark. If, for instance,
the kernel is approximated by one boson exchange,
i.e. ${\cal K}_a=g^2\theta(y-x)/(l^+-k^+)[(P-k)^--(l-k)^--(P-l)^-]$
and ${\cal K}_b=-g^2\theta(x-y)/(k^+-l^+)[k^--l^--(k-l)^-]$, where
$y=l^+/P^+$ and $x=k^+/P^+$, then one can easily see
${\cal K}_a(x,y)=-{\cal K}_b(y\to x,x\to y)$. Note also in this
simple one boson exchange case that there is only one light-front
time ordered diagram in each kernel and other contributions such
as $\theta(x-y)$-term in ${\cal K}_a$ and $\theta(y-x)$-term in
${\cal K}_b$ vanish since they contribute to pair creations from the
vacuum. Likewise, the vertex functions $\chi^{\rm f}$ and $\chi^g$
in Eq.~(\ref{had-pho1}) are the same as those in
Eq.~(\ref{had-pho2}) upon exchange of dummy variables
$(x,{\bf k}_\perp)\leftrightarrow (y,{\bf l}_\perp)$.

Exchanging $(x,{\bf k}_\perp)\leftrightarrow (y,{\bf l}_\perp)$
in Eq.~(\ref{had-pho2}), we now combine these two contributions,
${\cal F}^{nv}_{\pi(a)}$ and ${\cal F}^{nv}_{\pi(b)}$, to get
\bea\label{had-pho}
{\cal F}^{nv}_\pi(\zeta,x,t)&=& {\cal F}^{nv}_{\pi(a)}(\zeta,x,t)
+  {\cal F}^{nv}_{\pi(b)}(\zeta,x,t)\nonumber\\
&=&\int d^2{\bf k}_\perp
\int^1_\zeta dy\int d^2{\bf l}_\perp
\chi^{\rm f}(y,{\bf l}_\perp)
{\cal K}(x,{\bf k}_\perp; y,{\bf l}_\perp)
\chi^g(x,{\bf k}_\perp) S^+_{nv}(x,{\bf k}_\perp)
{\tilde\chi}^{\rm i}_a(x,{\bf k}_\perp)
\nonumber\\
&\times& \biggl[1
-\frac{S^+_{nv}(y,{\bf l}_\perp)}{S^+_{nv}(x,{\bf k}_\perp)}
\frac{{\tilde\chi}^{\rm i}_b(y,{\bf l}_\perp)}
{{\tilde\chi}^{\rm i}_a(x,{\bf k}_\perp)}\biggr],
\eea
where ${\cal K}={\cal K}_a$ and the relative($-$) sign between the
two contributions assures the removal of infrared singularity
at $x=y$ and ${\bf k}_\perp = {\bf l}_\perp$ (see also Ref.~\cite{Bur1}).
In Eq.~(\ref{had-pho}),
the $(x,{\bf k}_\perp)$ dependence in ${\cal F}^{nv}_{\pi(b)}$
comes solely from the term
${\cal K}(x,{\bf k}_\perp; y,{\bf l}_\perp)\chi^g(x,{\bf k}_\perp)$
(i.e. upper small loop in Fig.~\ref{Eff_photon}(b)),
which in principle includes the sum of all possible
intermediate mesons being coupled to the external current.

By putting the relative label of internal
momenta back into Eq.~(\ref{had-pho}), we obtain the same form of
${\cal F}^{nv}_\pi$ as that given by Eq.~(\ref{jnv})
\bea\label{jnv_new}
{\cal F}^{nv}_\pi(\zeta,x,t)
&=&\frac{-N_c}{2x(1-x)(1-x'')}\int
\frac{d^2{\bf k}_\perp}{16\pi^3}
\chi_{(2\to 2)}(x,{\bf k}_\perp)S^+_{nv}(x,{\bf k}_\perp)
\chi^g(x'',{\bf k''}_\perp)
\nonumber\\
&\times&\int^1_0\frac{dy'}{y'(1-y')}\int d^2{\bf l}_\perp
{\tilde{\cal K}}(x',{\bf k'}_\perp; y',{\bf l'}_\perp)
\chi'_{(2\to 2)}(y',{\bf l'}_\perp),
\eea
but now with
\bea\label{ktilde}
{\tilde{\cal K}}(x',{\bf k'}_\perp; y',{\bf l'}_\perp)
&\equiv& {\cal K}(x',{\bf k'}_\perp; y',{\bf l'}_\perp)
\biggl[
1-\frac{S^+_{nv}(y,{\bf l}_\perp){\tilde\chi}^{\rm i}_b(y,{\bf l}_\perp)}
{S^+_{nv}(x,{\bf k}_\perp){\tilde\chi}^{\rm i}_a(x,{\bf k}_\perp)}
\biggr].
\eea
Thus, in order to make our previous model~\cite{CJK_SPD} more complete,
the kernel ${\cal K}(x',{\bf k'}_\perp; y',{\bf l'}_\perp)$
given by Eq.~(\ref{jnv}) has to be replaced by
${\tilde{\cal K}}(x',{\bf k'}_\perp; y',{\bf l'}_\perp)$
which is free from any infrared singularity that might occur in the
kernel ${\cal K}$. Since it would be a formidable task to solve the kernel
${\tilde{\cal K}}$ directly, we follow the technique illustrated in our
previous analysis\cite{CJK_SPD} approximating $G_\pi$ defined by
\bea\label{New_Gpi}
G_\pi&\equiv&
\int^1_0\frac{dy'}{y'(1-y')}\int d^2{\bf l}_\perp
{\tilde{\cal K}}(x',{\bf k'}_\perp;
y',{\bf l'}_\perp) \chi_{(2\to2)}(y',{\bf l'}_\perp)
\eea
as a constant.
As we did in our previous analysis\cite{CJK_SPD}, we discuss the validity
of a constant $G_\pi$ approximation in the next section (Section IV)
where our numerical analysis is presented. Although the technical aspect
of our numerical analysis remains same, the replacement of kernel from
$\cal K$ to $\tilde{\cal K}$ amounts to the addition of new contribution
${\cal F}^{nv}_{\pi(b)}$ and thus effectively allow us to change the
gauge boson
wave function $\chi^g(x'',{\bf k''}_\perp)$ from the simple energy
denominator given by Eq.~(\ref{wfI}) to the LF vertex function identical
to the meson vertex function without introducing any additional parameters
\footnote{In principle, one
should use the vector(such as $\rho$) meson wave function replacing
the gauge boson vertex function.
We do not need to introduce any additional
parameters since we used the same gaussian wave function with the same
parameter $\beta$ for both $\pi$ and $\rho$ mesons
in our previous LFQM analysis \cite{CJ1}.}.

Comparing $\chi_{(2\to 2)}$ with our light-front
wave function given by Ref.~\cite{CJ1}, we identify
\be\label{LFvertex}
\chi_{(2\to 2)}(x,{\bf k}_\perp)=
\sqrt{\frac{8\pi^3}{N_c}}
\sqrt{\frac{\partial k_z}{\partial x}}
\frac{[x(1-x)]^{1/2}}{M_0}\phi(x,{\bf k}_\perp),
\ee
where the Jacobian of the variable tranformation
${\bf k}=(k_z,{\bf k}_\perp)\to
(x,{\bf k}_\perp)$ is obtained as $\partial k_z/\partial x=M_0/[4x(1-x)]$
and the radial wave function is given by
$\phi({\bf k}^2)=\sqrt{1/\pi^{3/2}\beta^3}\exp(-{\bf k}^2/2\beta^2)$,
which is normalized as $\int d^3k|\phi({\bf k}^2)|^2=1$. Note that
the radial wave function $\phi({\bf k}^2)$ is essentially the same as
the Brodsky-Huang-Lepage~\cite{BHL} LF wave function
$\phi^{\rm BHL}=\exp(-M^2_0 / 8\beta^2)$ up to a constant factor.

Thus, the effective gauge boson wave function
$\chi^g(x'',{\bf k''}_\perp)$  is given by
\be\label{LFGvertex}
\chi^g(x'',{\bf k''}_\perp)=
\sqrt{\frac{8\pi^3}{N_c}}
\sqrt{\frac{\partial k''_z}{\partial x''}}
\frac{[x''(1-x'')]^{1/2}}{M''_0}\phi^g(x'',{\bf k''}_\perp),
\ee
where $\partial k''_z/\partial x''=M''_{0}/[4x''(1-x'')]$.
Here, the radial wave function $\phi^g(x'',{\bf
k''}_\perp)$ is given by
\be\label{Gradial}
\phi^g({\bf k''}^2)=
\sqrt{\frac{1}{\pi^{3/2}\beta^3}}
\exp(-{\bf k''}^2/2\beta^2)\exp(\Delta^2/8\beta^2),
\ee
where ${\bf k''}^2=k''^2_z + {\bf k''}^2_\perp$ and
the virtuality $\Delta^2=t\neq 0$.

Substituting Eqs.~(\ref{LFvertex}) and~(\ref{LFGvertex}) into
Eqs.~(\ref{jv}) and~(\ref{jnv_new}), we obtain the valence and nonvalence
contributions to the GPDs of the pion in LFQM as follows
\bea\label{light-frontFv}
{\cal F}^{val}_\pi(\zeta,x,t)&=&
\int d^2{\bf k}_\perp
\sqrt{\frac{\partial k'_z}{\partial x'}}
\sqrt{\frac{\partial k_z}{\partial x}}
\phi'(x',{\bf k'}_\perp)
\phi(x,{\bf k}_\perp)\frac{({\bf k}_\perp\cdot{\bf k'}_\perp + m^2)}
{\sqrt{{\bf k}^2_\perp + m^2}\sqrt{{\bf k'}^2_\perp + m^2}},
\eea
and
\bea\label{light-frontFnv}
{\cal F}^{nv}_\pi(\zeta,x,t)
&=&-\sqrt{\frac{8\pi^3}{N_c}}\biggl(\frac{x''}{1-x'}\biggr)
\int d^2{\bf k}_\perp
\sqrt{\frac{\partial k_z}{\partial x}}
\sqrt{\frac{\partial k''_z}{\partial x''}}
\phi(x,{\bf k}_\perp)\phi^g(x'',{\bf k''}_\perp)
\nonumber\\
&\times&
\frac{{\bf k}_\perp\cdot{\bf k'}_\perp + m^2
+ x(1-x)x'(M^2-M^2_0)}
{ \sqrt{{\bf k}^2_\perp + m^2}
\sqrt{{\bf k''}^2_\perp + m^2} }
\int^1_0 dy'\int d^2{\bf l}_\perp
\sqrt{\frac{{\partial l'_z}}{{\partial y'}}}
\frac{{\tilde{{\cal K}}}(x',{\bf k'}_\perp; y',{\bf l'}_\perp)}
{\sqrt{{\bf l'}^2_\perp + m^2}}\phi'(y',{\bf l'}_\perp),
\eea
where Eq.~(\ref{light-frontFv}) is the same as our previous formula
Eq.~(28) in~\cite{CJK_SPD} but Eq.~(\ref{light-frontFnv}) is now
different from Eq.~(29) in~\cite{CJK_SPD} due to the modification
of the gauge boson vertex function.
It is worthwhile to note that the nonvalence
${\cal F}^{nv}_\pi$ in Eq.~(\ref{light-frontFnv}) receives not only
the on-mass shell propagating contribution (i.e. the term proportional
to $({\bf k}_\perp\cdot{\bf k'}_\perp + m^2)$) but also
the instantaneous contribution(i.e. the term proportional to
$x(1-x)x'(M^2-M^2_0)$) from the spectator quark
propagator. This also contrasts to the valence ${\cal F}^{val}_\pi$
given by Eq.~(\ref{light-frontFv}), which receives
only on-mass shell propagating contribution.
As we shall show in our numerical calculation, the
instantaneous contributions from the nonvalence ${\cal F}^{nv}_\pi$ becomes
substantial for large $\zeta>0.5$.

As discussed earlier, we shall treat the last $dy'd^2{\bf l}_\perp$-integral
term in Eq.~(\ref{light-frontFnv}) as a constant $G_\pi$ and check if indeed
a constant $G_\pi$ approximation is valid by varying the value of $\zeta$
in the sum rule expressed in terms of
${\cal F}^{val}_\pi$ and ${\cal F}^{nv}_\pi$, i.e.
\be\label{2sum}
F_\pi(t)=\int^1_0\frac{dx}{1-\frac{\zeta}{2}}{\cal F}_\pi(\zeta,x,t)
=\int^1_\zeta\frac{dx}{1-\frac{\zeta}{2}} {\cal F}^{val}_\pi(\zeta, x,t)
+ \int^\zeta_0 \frac{dx}{1-\frac{\zeta}{2}}{\cal F}^{nv}_\pi(\zeta, x,t),
\ee
for given $-t$. Varying the value of $\zeta$ in this sum rule,
Eq.~(\ref{2sum}), we can also check the frame-independence of our model
as we discuss in the next section(Section IV).

\subsection{The GPD value at the crossover: ${\cal F}_\pi (\zeta,\zeta,t)$}

Although the continuity of GPDs at $x=\zeta$ is assured in our formulation
presented in the last subsection (Section III B), the value of
${\cal F}_\pi (\zeta,\zeta, t)$ vanishes in our model calculation as
we shall see
in our numerical results (Section IV). The reason why this occurs in our
model calculation for the meson target is because the final state meson
(depicted in Fig.\ref{highFock}(b) and Fig.\ref{Eff_photon}) shares the
total longitudinal momentum fraction $1-\zeta$ only between the two
constituents so that, as $x-\zeta \rightarrow 0$ ({\it i.e.} the struck
constituent loses its longitudinal momentum completely), the single
spectator-constituent carries the total longitudinal momentum fraction
$1-\zeta$ and the final state two-body wavefunction at this kinematical point
($x=\zeta$) is zero. However, the situation is entirely
different for the three-body wavefunction such as in the proton target
because even at $x=\zeta$ the two spectator-constituents can still share
the total longitudinal momentum fraction $1-\zeta$ between themselves and
the final state three-body wavefunction does not vanish at $x=\zeta$. As
discussed in the literature\cite{KPV,FM} and mentioned in Section II,
the value of GPDs at $x=\zeta$ is directly related to the SSA that can be
measured in the scattering on an unpolarized target by the lepton
polarized parallel or antiparallel to its direction. The recent
measurements by HERMES and CLAS collaborations showed that the SSA is
indeed non-zero for the proton target and our observations based on the
three-body wavefunction are not inconsistent with these experimental
evidences for the non-zero GPDs of proton at $x=\zeta$.

By the same reasoning, if one includes the quark-antiquark-glue
three-body state (and any other multi-constituent-states) beyond the
ordinary two-body ($Q \bar{Q}$) constituents in our model calculation,
then it is in principle possible to get a non-zero
${\cal F}_{\pi}(\zeta,\zeta,t)$.
In the chiral-quark-soliton model\cite{PW}, it is shown that
${\cal F}_{\pi}(\zeta,\zeta,t)$ is non-zero.
Thus, if the experiment on the pion
is possible at all, then the SSA measurement would be crucial to
discreminate different models and also find how much percentage of
multi-constituent-component is necessary beyond our current $Q \bar{Q}$
state for the more realistic model construction.
However, it is interesting to note that our current model calculation
still satisfies the polynomiality conditions (See Eq.(\ref{sum-mom}))
and our result on the isosinglet GPD of the pion is qualitatively very
similar to the result obtained by the chiral-quark-soliton model satisfying
the soft pion theorem, as we comment in the following Section (Section IV).

\section{Numerical Results}

In our numerical calculations, we use the model parameters
$(m,\beta)=(0.22,0.3659)$ [GeV] obtained in Ref.~\cite{CJ1} for the
linear confining potential model.

\begin{figure}
\centerline{\psfig{figure=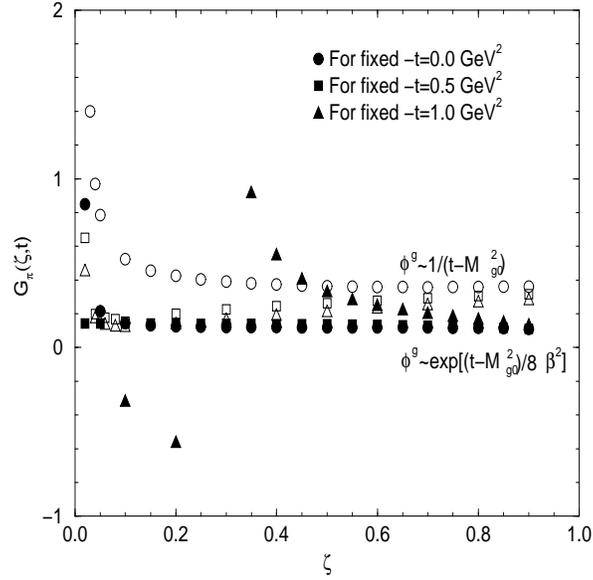,height=9cm,width=9cm}}
\caption{Comparison of the $\zeta$-dependence of $G_\pi$ between our new
gaussian type(black data) and previous monopole
type~\protect\cite{CJK_SPD}(white data) gauge boson wave functions
for different momentum transfers $-t=0$ (circle),
0.5(square), and 1 (triangle) [GeV$^2$],
respectively. The invariant mass of the $q\bar{q}$ pair at the gauge boson
vertex is represented as $M^2_{g0}=M''^2_0$.\label{Gfactor}}
\end{figure}

In Fig.~\ref{Gfactor}, we show the comparison of the
$\zeta$-dependence of $G_\pi$ between the two gauge boson wave functions,
i.e. our new gaussian type (black data),
$\phi^g\sim\exp[(t-M''^2_{0})/8\beta^2]$, and the previous~\cite{CJK_SPD}
monopole type (white data), $\phi^g\sim 1/(t-M''^2_{0})$ (see
Eq.(\ref{wfI})), for a few different momentum transfers,
$-t=0$ (circle), 0.5 (square), and  1.0 (triangle) [GeV$^2$], respectively.
The invariant mass of the $q\bar{q}$ pair at the gauge boson
vertex is represented as $M^2_{g0}=M''^2_0$ (see Eq.~(\ref{wfI}))
in the figure.
As one can see in Fig.~\ref{Gfactor}, both (black and white) $G_\pi$ values
show approximately constant behavior for small momentum transfer
$-t<1$ GeV$^2$ region.
It is not surprising to see that $G_\pi$ becomes very large
as $\zeta\to 0$, however, this does not cause a significant error in our
calculation because the nonvalence contribution in the very small $\zeta$
region is highly suppressed. Also, a rather large fluctuation of black
data for a rather large momentum transfer region $-t=1$ GeV$^2$
(even exhibiting a sign change near $\zeta=0.3$) is neither unexpected nor
troublesome because the nonvalence contribution is significantly
suppressed in this
large momentum transfer region where the gaussian wave function is
also exponentially reduced.
In order to check the reliability (i.e. frame-independence) of
our constant $G_\pi$ approximation, we compare
the exact solution of the pion electromagnetic form factor obtained
from $\zeta=0$ value with that of nonzero $\zeta$ value using our constant
(average) $G_\pi$ approximation.

\begin{figure}[t]
\centerline{\psfig{figure=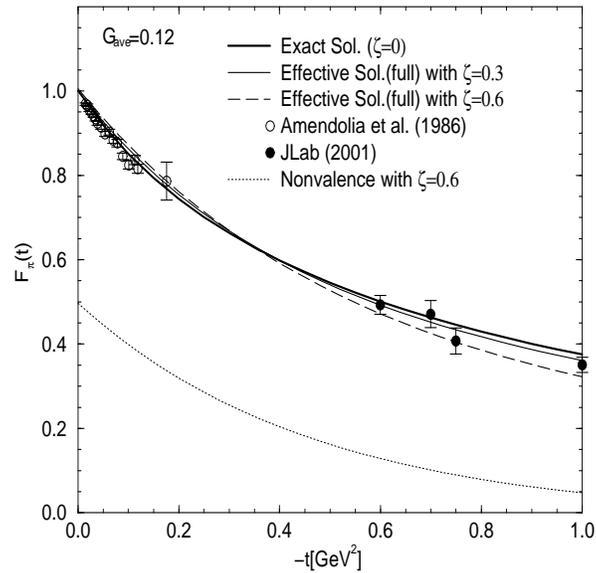,height=9cm,width=9cm}}
\caption{The effective solution of pion form factor
using a common average $G_\pi=G_{\rm ave.}=0.12$ value with
gaussian type gauge boson wave function
for $\zeta=0.3$ (thin solid line) 0.6 (long-dashed line)
compared with the exact solution (thick solid line) as well as experimental
data~\protect\cite{Amen,JLab}. The dotted line represents
the nonvalence contributions to the form factor for $\zeta=0.6$
case.  \label{Piave}}
\end{figure}

In Fig.~\ref{Piave}, we show our effective solutions of the pion
form factor with gaussian $\phi^g(x'',{\bf k''}_\perp)$
for $\zeta=0.3$ (thin solid line) and 0.6 (long-dashed line)
cases obtained from our average value of $G_\pi=G_{\rm ave.}=0.12$
and compare them with the exact solution (thick solid line) as well
as the experimental data~\cite{Amen,JLab}. The dotted line
represents the nonvalence contributions to the form
factor for $\zeta=0.6$ case, which increase (decrease) as $\zeta$ gets
larger (smaller). Also, there are
$-t_{\rm min}$ values for nonzero $\zeta$ due to
$\Delta^2_\perp\geq 0$ (see Eq.~(\ref{del2})).
We thus use the analytic continuation by
changing $\Delta_\perp$ to $i\Delta_\perp$ in
Eqs.~(\ref{light-frontFv}) and~(\ref{light-frontFnv}) to
obtain the result for $0\leq -t\leq -t_{\rm min}$ where there is
no singularity.  A continuous behavior of the form factor near
$-t_{\rm min}$ confirms the analyticity of our model calculation.
As far as the $G_\pi$ and the form factor calculations
are concerned, the results obtained from the gaussian type gauge boson
wave function are overall not much different from our previous
results~\cite{CJK_SPD} obtained from the monopole type gauge boson
wave function.  However, as we shall see below, they are distinguished
by the calculation of the GPDs.

\begin{figure}[t]
\psfig{figure=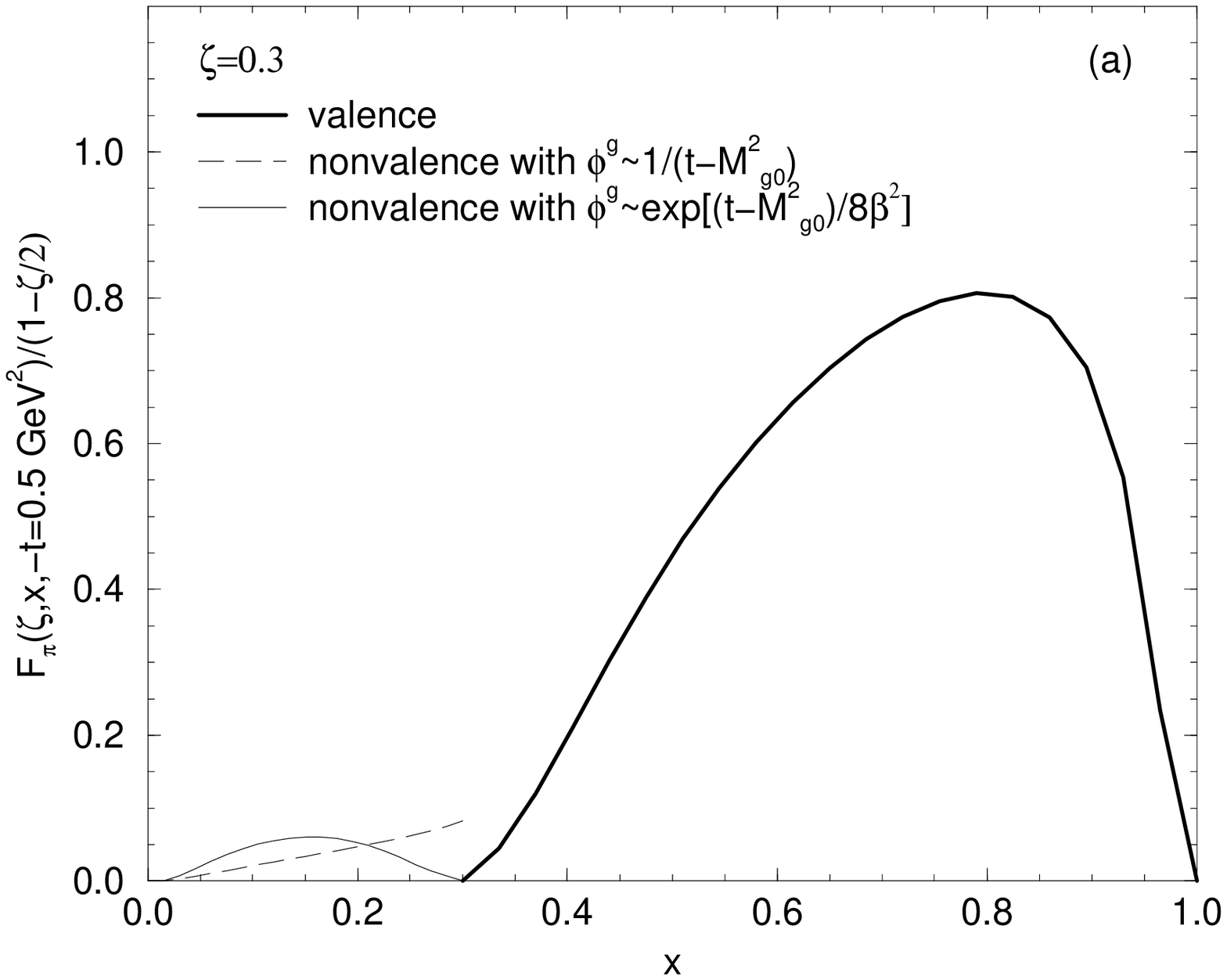,height=10cm,width=8.5cm}
\psfig{figure=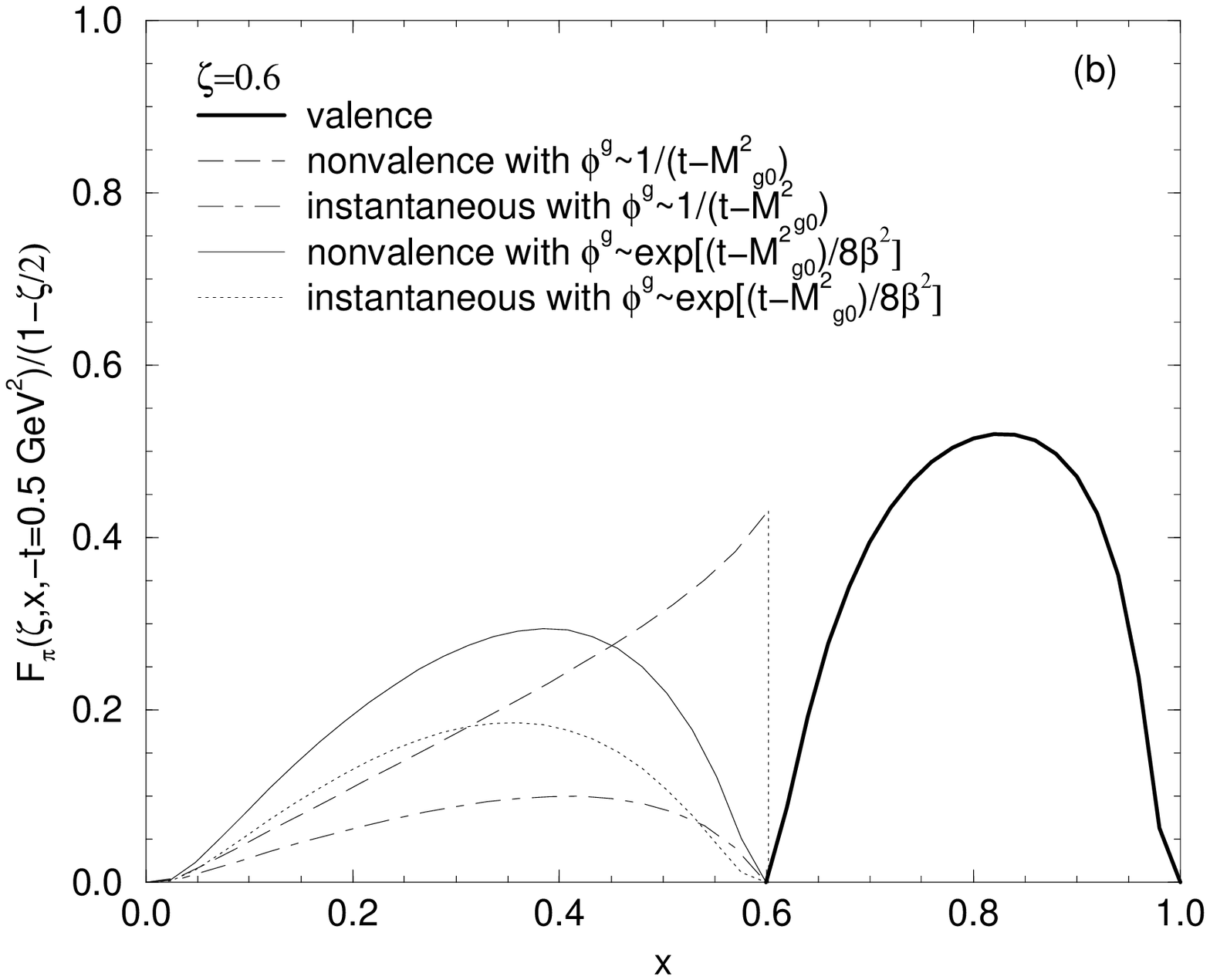,height=10cm,width=8.5cm}
\caption{Generalized parton distributions of the pion at
$-t=0.5$ GeV$^2$ with $\zeta=0.3$ in (a), and 0.6 in (b),
respectively. \label{SQD01}}
\end{figure}

In Fig.~\ref{SQD01}, we compare the nonvalence contributions(in ERBL region)
to the GPDs ${\cal F}_\pi(\zeta,x,t)/(1-\zeta/2)$
of the pion obtained from the gaussian type gauge boson wave function(thin
solid line) with those obtained from the monopole type wave
function(long-dashed line) for fixed momentum transfer
$-t=0.5$ GeV$^2$ but with different skewedness parameters
$\zeta=0.3$ in (a) and 0.6 in (b), respectively.
The thick solid line represents the valence contribution(in DGLAP region),
which is common
for both gaussian and monopole type gauge boson wave functions.
Our results are obtained from average values of $G_\pi$, i.e.
$G_\pi=0.12$ for the gaussian type gauge boson wave function
and 0.32~\cite{CJK_SPD} for the monopole type gauge boson wave function.
We also plot the instantaneous contributions to the GPDs with the gaussian
type(dotted line) and the monopole type(dot-dashed line) gauge boson
wave functions for $\zeta=0.6$ case.
We note that the results of the GPDs shown in
Fig.~\ref{SQD01} with the average $G_\pi$ values for both
gaussian and monopole type gauge boson wave function cases  are very
close to the exact ones, i.e. frame-independent, as one can deduce
from Fig.~\ref{Piave}
for the gaussian wave function and Fig.~9 in~\cite{CJK_SPD} for the
monopole type wave function, respectively.

As shown in Fig.~\ref{SQD01}, the most dramatic change in our updated
result of the pion GPDs is that the solutions with the gaussian type
gauge boson wave function show the continuity at the crossover($x=\zeta$),
while the solutions with the monopole type\cite{CJK_SPD}
gauge boson wave function show discontinuity at the crossover.
As we discussed before,
such discontinuity at $x=\zeta$ with the monopole type gauge boson
wave function is just an artifact due to
the different $x\rightarrow\zeta$ behavior between the gauge boson
vertex ($\chi^g$) and the hadronic vertex ($\chi_{(2 \to 2)})$ functions
(see Eq.(\ref{wfI})).
This discontinuity problem of the GPDs appeared
in our previous analysis~\cite{CJK_SPD} is now resolved by realizing that
the same type of bound state wave function should be used for both hadron
and gauge boson.
Furthermore, we can obtain finite results for the scattering amplitudes
for the DVCS region given by Eq.~(\ref{CDVCS2})
since both valence and nonvalence solutions of the GPDs are
not only continuous but also vanish at the crossover($x=\zeta$)
as well as at $x=0$.
As we discussed in Section III C, ${\cal F}_{\pi}(\zeta,\zeta,t)$ can be in
principle non-zero if the multi-constituent-states are included beyond
the ordinary $Q \bar{Q}$ state. Within our current $Q \bar{Q}$ model,
however, we have checked the $n$th moment $F_n (\zeta,t)$ in
Eq.(\ref{sum-mom}) for up to $n=3$ as well as the isosinglet GPDs.
With our updated formulation presented in Section III B, we did not find
any difference from our previous results, Figs. 10 and 11 of
Ref.\cite{CJK_SPD}, based on the formulation summarized in Section III A
except the change of $G_\pi$ value discussed above. Thus, we do not
show those figures in duplication but have confirmed that the
polynomiality conditions are satisfied in our current model and our
result on the isosinglet GPD of the pion is qualitatively very
similar to the result (Fig.5 of Ref.\cite{PW}) obtained by the
chiral-quark-soliton model including the D-term in the ERBL region.
The D-term generates the highest power in the polynomial, {\it e.g.}
the second moment of the pion isosinglet GPD is given by
$F_2^{I=0}(\zeta,t=0) = \frac{1}{2}(1+C\zeta^2)$, where $C = - 1/4$ in
the chiral limit\cite{PW}\footnote{The skewedness parameter $\xi$ in 
Ref.\cite{PW} is related to $\zeta$ by $\xi = {\zeta \over 
{1-\frac{\zeta}{2}}}$. Note, however, that the difference between $\xi$ 
and $\zeta$ is negligible in the limit $t \rightarrow 0$ because the 
physical region of $\zeta$ is restricted by Eq.(\ref{range}).}
Our model calculation from Figs. 10 and 11 of
Ref.\cite{CJK_SPD} gives $C = -0.2843$. Incidentally, we note from Fig.5
of Ref.\cite{PW} that the chiral-quark-soliton model not neglecting the
dynamical quark mass generated in the spontaneous breaking of chiral
symmetry also gives the value of $|C|$ slightly greater than 1/4.

\begin{figure}[t]
\psfig{figure=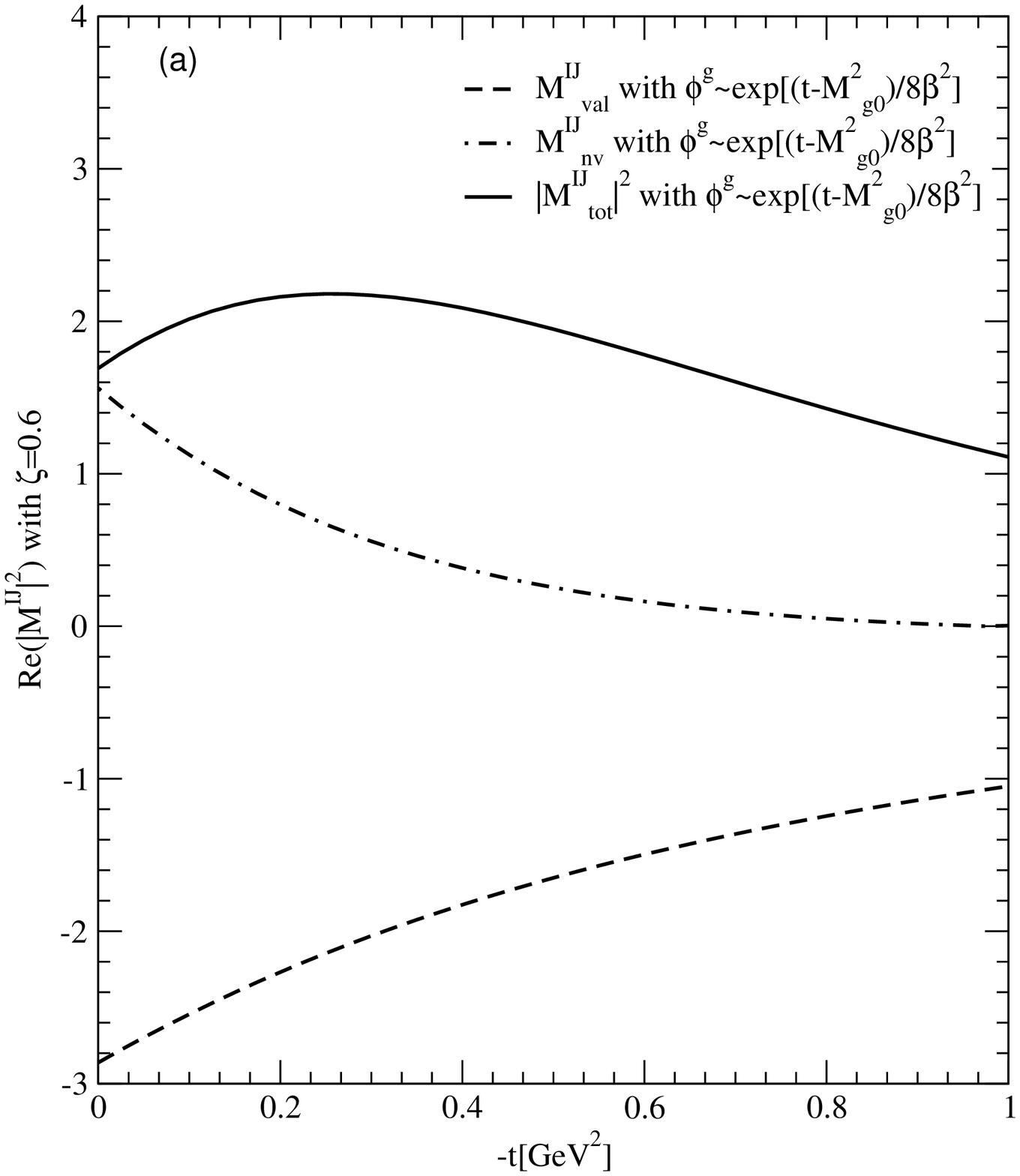,height=10cm,width=8.5cm}
\psfig{figure=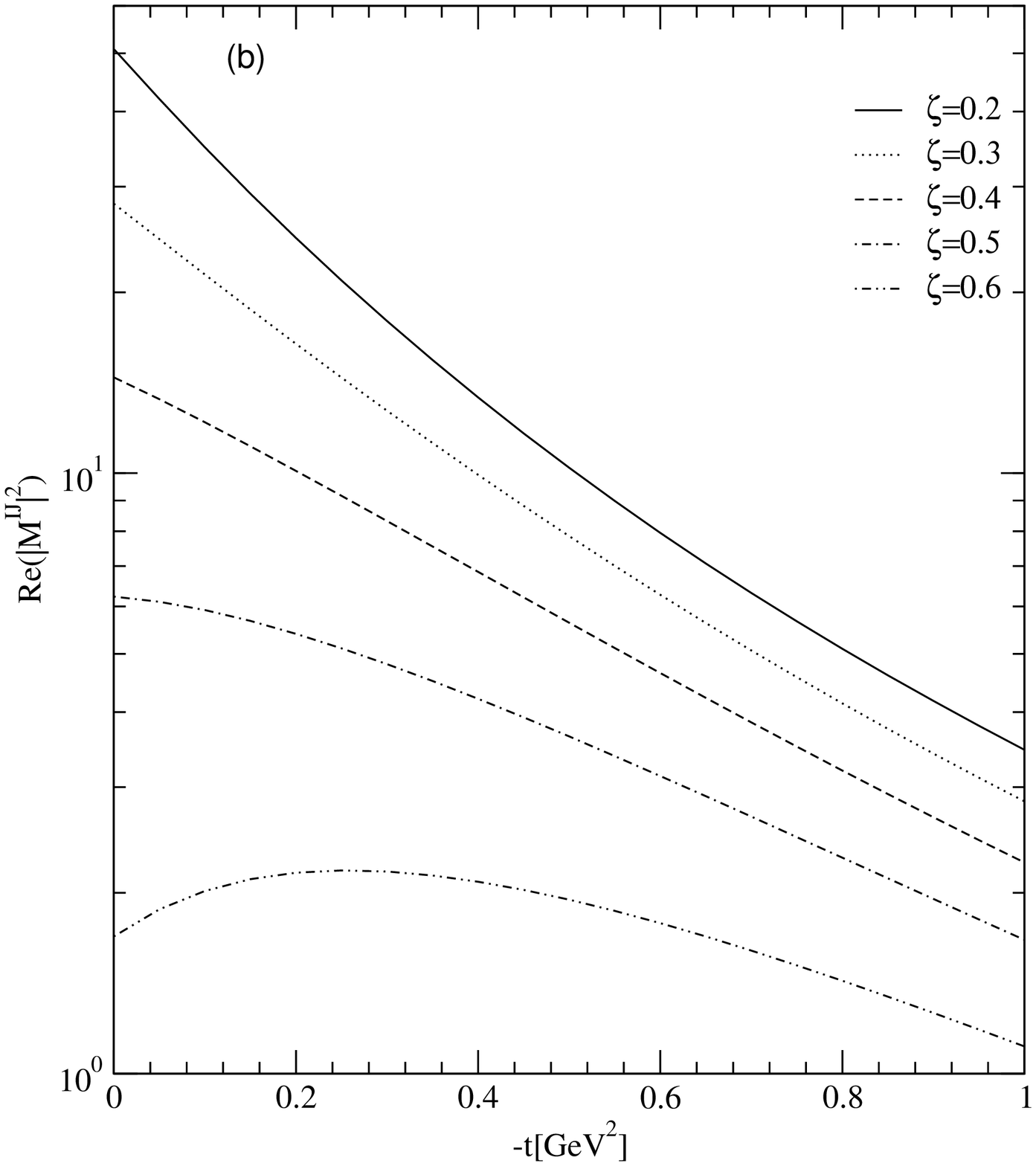,height=10cm,width=8.5cm}
\caption{ (a) The real part of the DVCS amplitude
$M^{IJ}=M^{\uparrow\uparrow}+M^{\downarrow\downarrow}$(see
Eq.~\protect(\ref{CDVCS2})) obtained from the
gaussian type gauge boson wave function and the average $G_\pi=0.12$ for
fixed $\zeta=0.6$:
The thick solid, dashed, dot-dashed lines represent the total squared
amplitude Re$(M^{IJ})^2$, valence part of $M^{IJ}=M^{IJ}_{val}$, and
nonvalence part of $M^{IJ}=M^{IJ}_{nv}$, respectively.
(b) The $\zeta$ dependence of the DVCS squared amplitude
Re($|M^{IJ}|^2$). \label{ReM}}
\end{figure}

In Fig.~\ref{ReM}(a), we show the real part of the DVCS
squared amplitude (thick solid line),
Re($|M^{IJ}|^2$)=Re($|M^{\uparrow\uparrow}+M^{\downarrow\downarrow}|^2$)
=4Re($|M^{\uparrow\uparrow}|^2$) in Eq.~(\ref{CDVCS2}),
obtained from the gaussian type gauge boson wave function
and the average $G_\pi=0.12$
for the range of  $0\leq -t\leq 1$ [GeV$^2$] with fixed $\zeta=0.6$.
The thick dashed and dot-dashed lines represent the valence
($M^{IJ}=M^{IJ}_{val}$) and nonvalence ($M^{IJ}=M^{IJ}_{nv}$) contributions
to the total amplitude, respectively.
It is interesting to note that the valence and
nonvalence amplitudes interfere with each other destructively
not constructively. We also note that the DVCS squared amplitude
Re($|M^{IJ}|^2$) for the monopole type gauge boson wave function
shows the logarithmic divergent behavior as expected due to the
discontinuity at the crossover $x=\zeta$. We thus do not present the
corresponding result for the monopole type gauge boson wave function.
In Fig.~\ref{ReM}(b), we show the $\zeta$ dependence of Re($|M^{IJ}|^2$)
for the gaussian type gauge boson wave function.
In our model calculations, the DVCS squared amplitude increases as $\zeta$
decreases.
As in the case of Fig.~\ref{Piave}, we confirm the analyticity of our
calculation from the continuous behavior of our results near $-t =
-t_{\rm min}$.

\section{Conclusion}
In this work, we reinvestigated our previous light-front quark model
analysis~\cite{CJK_SPD} of
the GPDs in the deeply virtual Compoton scattering (DVCS) of the pion,
$\gamma^* \pi\to\gamma \pi$.  We improved our previous
effective treatment~\cite{CJK_SPD} of handling
nonvalence (or higher Fock-states) contributions
to the GPDs with the inclusion of the BS kernel for the $q\bar{q}$ pair
annihilating into the gauge boson(see also~\cite{Ein,Bur1,TM} for a similar
discussion) in addition to that for the $q\bar{q}$ pair merging
into the final state
meson(see Fig.\ref{Eff_photon}). It leads us to have a vector meson
dominance picture at the gauge boson vertex within our constant
approximation of $G_\pi$ in Eq.~(\ref{New_Gpi}) and in turn
to use the same type of gaussian wave function
at both gauge boson and meson vertices.
The continuity of the GPDs at the
crossover($x=\zeta$) between the DGLAP and ERBL regions is
ensured by a consistent treatment of the gauge boson and meson
wave functions. Subsequently, the real part of the
DVCS amplitude $M^{IJ}$ is calculated and its finiteness is confirmed
(see Figs.\ref{ReM}(a) and (b)).
In our current $Q \bar{Q}$ constituent model, we obtain
${\cal F}_{\pi}(\zeta,\zeta,t) = 0$ indicating that the single spectator
cannot share its longitudinal momentum at $x=\zeta$. In principle,
${\cal F}_{\pi}(\zeta,\zeta,t)$ can be non-zero if our model
is extended to include
the multi-constituent-components. However, our current model is capable
of satisfying the polynomiality conditions and yields qualitatively very
similar results on the isosinglet GPD of the pion indicating that the
D-term is effectively included in the ERBL region. Also, one should note
that our model is not inconsistent with the recent HERMES and CLAS data
on the SSA for the proton target.

\begin{figure}
\centerline{\psfig{figure=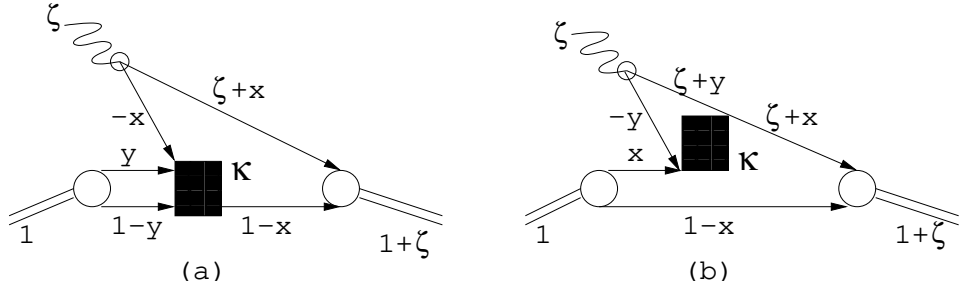,height=1.5in,width=5in}}
\caption{Illustration of time reversal($\Delta\to-\Delta$)
process of the GPDs in Fig.~\protect\ref{Eff_photon}.
\label{Eff_photon_TS}}
\end{figure}

Finally, the GPD of the pion is invariant under the
time reversal symmetry($\Delta\to-\Delta$)\footnote{Although it
might be convenient to choose a ``symmetric" light-front
frame~\cite{XJ1} for the momenta of the initial and final target
hadrons to see $\Delta\to-\Delta$ symmetry, the physical
interpretation of the initial and final hadron states in terms of
light-front wave function is more clear in the present asymmetric
frame than in the symmetric one.}. As illustrated schematically
in Fig.~\ref{Eff_photon_TS}, it is not difficult to see that
our effective method does not violate the time reversal symmetry.
Using our model calculation of the DVCS amplitude in which the final
photon is emitted by the pion, one can calculate the differential cross
section of the virtual Compton scattering of the pion, i.e. $\pi e\to\pi
e\gamma$, by incorporating the Bethe-Heitler(BH) process~\cite{DG,UO} in
which the final photon is emitted by either the incoming electron or the
outgoing electron. Considerations along this line are in progress.
\begin{center}
{\large\bf Acknowledgements}
\end{center}
The work of HMC and LSK was supported in part by the NSF
grant PHY-00070888 and that of CRJ by the US DOE under grant
No. DE-FG02-96ER40947. The North Carolina Supercomputing Center and
the National Energy Research Scientific Computer Center are also
acknowledged for the grant of Cray time.

\end{document}